\documentclass[aps,twocolumn,prd,epsf,showpacs,amsmath,amssymb]{revtex4}

\usepackage{graphicx}
\usepackage{dcolumn}
\usepackage{bm}

\newcommand{\be}{\begin{equation}}
\newcommand{\ee}{\end{equation}}
\newcommand{\ba}{\begin{eqnarray}}
\newcommand{\ea}{\end{eqnarray}}
\newcommand{\ban}{\begin{eqnarray*}}
\newcommand{\ean}{\end{eqnarray*}}

\newcommand{\n}[1]{\label{#1}}

\newcommand{\non}{\nonumber}
\newcommand{\eq}[1]{(\ref{#1})}

\newcommand{\hh}{\, ,\hspace{0.5cm}}
\newcommand{\hhh}{\, ,\hspace{0.2cm}}

\newcommand{\T}{\ensuremath{\theta}}
\newcommand{\al}{\ensuremath{\alpha}}

\newcommand{\ph}{\ensuremath{\phi}}
\newcommand{\x}{\ensuremath{\zeta}}
\newcommand{\E}{{\mathbb E}}
\newcommand{\C}{{\mathbb C}}
\newcommand{\R}{{\mathbb R}}
\newcommand{\pht}{\ensuremath{\tilde{\phi}}}
\newcommand{\tpsi}{\ensuremath{\tilde{\psi}}}

\newcommand{\pa}{\partial}

\begin{document}

\title{Surface Geometry of 5D Black Holes and Black Rings}      

\author{Valeri P. Frolov and Rituparno Goswami}
\affiliation{Theoretical Physics Institute, University of Alberta,
Edmonton, Alberta, Canada, T6G 2G7}

\email{frolov@phys.ualberta.ca}

\email{goswami@phys.ualberta.ca}

\date{\today}

\begin{abstract}    We discuss geometrical properties of the horizon
surface of  five-dimensional rotating black holes and black rings. 
Geometrical invariants characterizing these 3D geometries are
calculated.  We obtain a global embedding  of the 5D rotating black
horizon surface into a flat space. We also describe  the
Kaluza-Klein reduction of the black ring solution (along the direction
of its rotation) which relates this solution to the 4D metric
of a static black hole distorted by the presence of external scalar
(dilaton) and vector (`electromagnetic') field. The properties of the
reduced black hole horizon and its embedding in $\E^3$ are briefly
discussed.
\end{abstract}

\pacs{04.70.Bw, 04.50.+h, 02.40.Ky \hfill  
Alberta-Thy-17-06} 

\maketitle

\section{Introduction}

Black objects (holes, strings, rings etc.) in higher dimensional
spacetimes have attracted a lot of attention recently. The existence
of higher than 4 dimensions of the spacetime is a natural consequence
of the consistency requirement in the string theory. Models with
large extra dimensions, originally proposed to solve such
longstanding fundamental `puzzles' as the hierarchy and cosmological
constant problems, became very popular recently. In these models mini
black holes and other black objects play a special role serving as
natural probes of extra dimensions. This is one of the reasons why
the questions what kind of black objects can exist in higher
dimensions and what are their properties are now discussed so
intensively. 

Higher dimensional generalizations of the Kerr metric for a rotating
black holes were obtained quite long time ago by Myers and
Perry~\cite{mp1}. In a $D$-dimensional spacetime the MP metrics
besides the mass $M$ contain also $[(D-1)/2]$  parameters connected
with the independent components of the angular momentum of the black
hole. (Here $[A]$ means the integer part of $A$.) The event horizon
of the MP black holes has the spherical topology $S^{D-2}$.  This
makes them in many aspects similar to the 4D Kerr black hole.
According to the Hawking theorem~\cite{Hawk} any  stationary black
hole in a 4D spacetime obeying the dominant energy condition has the
topology of the horizon $S^2$. Black hole surface topologies distinct
from $S^2$ are possible if the dominant energy condition is violated
\cite{GeHa:82}. Moreover, a vacuum stationary black is uniquely
specified by its mass and angular momentum.  Recent discover of black
ring solutions ~\cite{br1}-\cite{br3} demonstrated that both the
restriction on the topology of the horizon and the uniqueness property
of black holes are violated in the 5D spacetime.

In this paper we discuss the geometry of the horizon surfaces of 5D
black rings  and a 5D black holes with one rotation parameter. A
similar problem for the 4D rotating black  holes was studied in
detail by Smarr \cite{kerr1}. We generalize his approach to the 5D
case. After a brief summary of known properties of 3D round spheres and
tori in the flat 4D space (Section~2) we  consider a geometry of 3D
space which admits 2 orthogonal commuting Killing vectors  (Section~3). In
particular we calculate its Gauss curvature. In Section~4
we apply these results to the horizon surface of 5D rotating black
hole with one rotation parameter. The embedding of this 3D surface
into the flat spacetime is considered in Section~5. The horizon
surface geometry for a 5D rotating black ring is discussed in
Section~6. This Section considers also a Kaluza-Klein reduction of
the black ring metric along the direction of its rotation which maps
this solution onto a black hole solution of 4D Einstein equations with the
dilaton and `electromagnetic' fields. The geometry and embedding of
the horizon in the $\E^3$ for this metric is obtained. Section~7
contains the discussion of the results.

\section{Sphere $S^3$ and torus $S^2\times S^1$ in $\E^4$}

\subsection{Sphere $S^3$}

In this section we briefly remind some known properties of  a 3D
sphere and a torus in a flat 4D space. 

Consider 4-dimensional Euclidean space $\E^4$ and denote by $X_i$
($i=1,\ldots,4$) the Cartesian coordinates in it.   A 3-sphere
consists of all points equidistant from a single point $X_i=0$ in
$\R^4$. A unit round sphere $S^3$ is a surface defined by the
equation $\sum_{i=1}^4 X_i^2=1$. Using complex coordinates
$z_1=X_1+iX_2$ and $z_2=X_3+iX_4$ one can also equivalently define
the unit 3-sphere as a subset of  $\C^2$ 
\begin{equation}
S^3=\left\{(z_1,z_2)\in\C^2|\;|z_1|^2+|z_2|^2=1\right\} \, .
\label{eq:s31}
\end{equation}
We use the embedding of $S^3$ in $\C^2$  to introduce the 
{\em Hopf} co-ordinates ($\theta,\phi,\psi$) as,
\begin{equation}
z_1=\sin(\theta)e^{i\phi}\;\;;\;\;z_2=\cos(\theta)e^{i\psi} \, .
\label{eq:s32}
\end{equation}
Here $\T$ runs over the range $[0,\pi/2]$, and $\phi$ and $\psi$ 
can take any values between $0$ and $2\pi$. In these co-ordinates 
the metric on the 3-sphere is 
\begin{equation}
ds^2=d\T^2+\sin^2\T d\phi^2+\cos^2\T d\psi^2 \, .
\label{eq:s33}
\end{equation}
The volume of the unit 3-sphere is $2\pi^2$. Coordinate lines of
$\phi$ and $\psi$ are circles. The length of these circles take the
maximum value $2\pi$ at $\theta=\pi/2$ for $\phi$-line and at
$\theta=0$ for $\psi$-line, respectively. These largest circles are
geodesics. Similarly, the coordinate lines of $\theta$ coordinate are
geodesics. For the fixed values of $\phi=\phi_0$ and $\psi=\psi_0$
and $\theta\in [0,\pi/2]$ this line is a segment of the length $\pi/2$
connecting the fixed points of the Killing vectors $\partial_{\phi}$ and
$\partial_{\psi}$. Four such segments $\phi=\phi_0,\phi_0+\pi$,
$\psi=\psi_0,\psi_0+\pi$ form the largest circle of the length
$2\pi$.

The surfaces of constant $\T$ are flat {\em tori} $T^2$. For
instance,  $\T=\T_0$ can be cut apart to give a rectangle with
horizontal edge length $\cos\T_0$  and vertical edge length
$\sin\T_0$. These tori are called {\em Hopf tori} and  they are
pairwise linked. The  fixed points  of the vectors $\pa_{\phi}$ and
$\pa_{\psi}$ ($\T=0$ for $\pa_{\phi}$
and $\T=\pi/2$ for $\pa_{\psi}$) form a pair of linked great
circles. Every other Hopf torus passes between  these circles.  The
equatorial Hopf torus is the one which can be made from a  square.
The others are all rectangular. Also we can easily see that   the
surfaces of constant $\phi$ or constant $\psi$ are {\em half}
2-spheres or topologically disks.

\subsection{Torus ${\cal T}^3=S^2\times S^1$}

The equation of a torus ${\cal T}^3=S^2\times S^1$ in $\E^4$ is
\be
X_1^2+X_2^2+(\sqrt{X_3^2+X_4^2}-a)^2=b^2\, .
\ee
The surface ${\cal T}^3$ is obtained by the rotation of a sphere $S^2$
of the radius $b$ around a circle $S^1$ of the radius $a$ ($a>b$).
Let us define toroidal coordinates as
\ba
X_1&=&{\alpha\sin \hat{\theta}\over B}\cos\phi \hhh X_2={\alpha\sin
\hat{\theta}\over B}\sin\phi\, ,\non\\
X_3&=&{\alpha\sinh \eta\over B}\cos\psi \hhh X_4={\alpha\sinh \eta\over
B}\sin\psi \, ,
\ea
where $B=\cosh \eta -\cos\hat{\theta}$. The toroidal coordinates
$(\eta,\hat{\theta},\phi,\psi)$ change in the following intervals
\be
0<\eta<\infty\hhh
0\le \hat{\theta}\le \pi\hhh
0\le \phi, \psi \le 2\pi \, .
\ee
The flat metric in this coordinates takes the form
\be\n{mtor}
ds^2={\alpha^2\over B^2}\left( d\eta^2+\sinh^2\eta d\psi^2
+d\hat{\theta}^2+\sin^2\hat{\theta} d\phi^2\right)\, .
\ee
In these coordinates the surface of constant $\eta=\eta_0$ is a torus
${\cal T}^3$ and one has
\be
\alpha=\sqrt{a^2-b^2}\hh
\cosh \eta_0=a/b\, .
\ee
Introducing new coordinates $y=\cosh \eta$ ans $x=\cos \hat{\theta}$ one
can also write the metric
\eq{mtor} in the form \cite{br3}
\ba
ds^2&=&{\alpha^2\over (y-x)^2}\left[ {dy^2\over y^2-1}+(y^2-1)d\psi^2 \right.\non \\
&+&\left. {dx^2\over 1-x^2}+(1-x^2)d\phi^2\right]\, .
\ea

The points with $\eta<\eta_0$ lie in the exterior of ${\cal T}^3$. 
The induced geometry on the 3-surface $\eta=\eta_0$ is
\be\n{cantor}
ds^2={a^2-b^2\over (a-b\cos\hat{\theta})^2} [(a^2-b^2)
d\psi^2+b^2(d\hat{\theta}^2+\sin^2\hat{\theta} d\phi^2)]\, .
\ee
This metric has 2 Killing vectors, $\pa_{\phi}$ and $\pa_{\psi}$. The
first one has 2 sets of fixed points, $\theta=0$ and $\theta=\pi$,
which are circles $S^1$. The second Killing vector, $\pa_{\psi}$ does
not have fixed points.  The 3-volume of the torus ${\cal T}^3$ is
$8\pi^2 a b^2$.

Since the sections $\psi=$const are round spheres, instead of
$\hat{\theta}$ it is convenient to use another coordinate,
$\theta\in[0,\pi]$, 
\be
\sin\theta={\sqrt{a^2-b^2} \sin\hat{\theta} \over
a-b\cos\hat{\theta}}\, .
\ee
Using this coordinate one can rewrite the metric \eq{cantor} in the
form
\be\n{mt}
ds^2=(a+b\cos\theta)^2 d\psi^2+b^2(d{\theta}^2+\sin^2{\theta} d\phi^2)\, .
\ee
Once again we can easily see
that  the surfaces of constant $\T$ are flat {\em tori} $T^2$ except
for  $\T=0$ or $\T=\pi$, which are circles. The surfaces of constant
$\psi$ are  2-spheres whereas the surfaces of constant $\phi$ are
2-tori. 

Sometimes it is convenient to consider special foliations of ${\cal
T}^3$ \cite{fol}. This foliation is a kind of "clothing" worn on a
manifold,  cut from a stripy fabric. These stripes are called plaques
of the foliation.  On each sufficiently small piece of the manifold, 
these stripes give the manifold a local product structure.  This
product structure does not have to be consistent outside local
patches,  a stripe followed around long enough might return to a
different, nearby stripe. As an example for foliations let us
consider the manifold $\R^3$. The foliations are generated by two
dimensional leaves or plaques with one co-ordinate as constant. That
is, the surfaces $z=constant$, would be the plaques of the 
foliations and in this case there is a global product structure.  
Similarly one can consider the foliations of $S^2\times S^1$. Fig.(1)
shows the transverse {\em Reeb} foliations of the cylindrical section
of $S^2\times S^1$ \cite{fol}. We can see the the stacking of
spherical shaped  plaques giving rise to the cylindrical section. 
\begin{figure}[t]
\begin{center} 
\includegraphics[height=4.5cm,width=7cm]{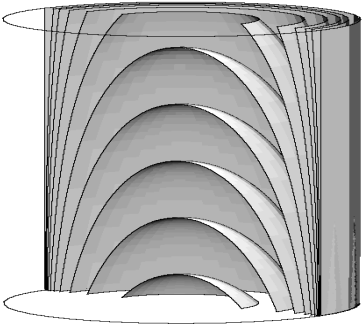} 
\caption{This picture shows the transverse Reeb foliations of the
cylindrical section  of $S^2{\times}S^1$. The two dimensional
spherical shaped stripes or `plaques' are stacked giving  rise two a
cylindrical section of the 3-torus. (courtesy:
http://kyokan.ms.u-tokyo.ac.jp)} \label{f1} 
\end{center} 
\end{figure}

\section{Geometry of 3-dimensional space with 2 orthogonal commuting
Killing vectors}

As we shall see both  metrics of the horizon surface of a 5D black
ring and  a  black hole with one rotation parameter can be written in
the form 
\begin{equation}
ds_H^2=f(\x)d\x^2+g(\x)d\phi^2+h(\x)d\psi^2\, .
\label{eq:genmetric}
\end{equation}
Here $f$, $g$ and $h$ are non-negative functions of the co-ordinate 
$\x$. One can use an ambiguity in the choice of the coordinate $\x$
to put $f=1$. For this choice $\x$ has the meaning of the proper
distance along $\x$-coordinate line. We call such a parametrization
canonical. The co-ordinates $\phi$ and $\psi$ have a period of $2\pi$
and  $\x\in[\x_{0},\x_{1}]$.  $\partial_{\phi}$ and $\partial_{\psi}$
are two mutually orthogonal Killing vectors. If $g(\x)$ ($h(\x)$)
vanishes at some point then the Killing vector $\partial_{\phi}$
($\partial_{\psi}$) has a fixed point at this point. The metric
\eq{eq:genmetric} does not have a cone-like singularity at a fixed
point of $\partial_{\phi}$ if at this point the following condition
is satisfied
\be\n{cone}
{1\over 2\sqrt{hf}}{dh\over d\x}=1\, .
\ee
A condition of regularity of a fixed point of $\partial_{\psi}$ can be
obtained from \eq{cone} by changing $h$ to $g$.

By comparing the metric \eq{eq:genmetric} with the metric for the
3-sphere \eq{eq:s33} one can conclude that \eq{eq:genmetric} describes
the geometric of a distorted 3D sphere if $g$ and $h$ are positive
inside some interval $(\x_1,\x_2)$, while $g$ vanishes at one of its end
point (say $\x_1$) while $h$ vanishes at the other (say $\x_2$).
Similarly, by comparison \eq{eq:genmetric} with \eq{mt} one concludes
that if for example $g$ is positive in the interval $(\x_1,\x_2)$ and
vanishes at its ends, while $h$ is positive everywhere on this interval,
including its ends, the metric \eq{eq:genmetric} describe a topological
torus.

For the metric (\ref{eq:genmetric}), the non-vanishing components of the 
curvature tensor are,
\ba
R_{\x\phi\x\phi}&=&\frac{g'(fg)'}{4fg}-\frac{1}{2}g''\, ,
\label{eq:riemann1}\\
R_{\x\psi\x\psi}&=&\frac{h'(fh)'}{4fh}-\frac{1}{2}h''\, ,
\label{eq:riemann2}\\
R_{\phi\psi\phi\psi}&=&-\frac{g'h'}{4f}  \, .     
\label{eq:riemann3}
\ea
Here ($'$) denotes the differentiation with respect to co-ordinate $\x$.

Denote by $e^{i}_{{a}}$ ($i, {a}=1,2,3$) 3 orthonormal
vectors and introduce the Gauss curvature tensor as follows
\be
K_{ab}=-R_{ijkl}e^{i}_{{a}}e^{j}_{{b}}e^{k}_{{a}}e^{l}_{{b}}\,
.
\ee
The component $K_{ab}$ of this tensor coincides with the curvature in
the 2D direction for the 2D plane spanned by $e^{i}_{{a}}$ and
$e^{j}_{{b}}$. One has
\be\n{KK}
\sum_{b=1}^3 K_{ab}=R_{ij}e^{i}_{{a}}e^{j}_{{a}}\hh
\sum_{a=1}^3\sum_{b=1}^3 K_{ab}=R\, .
\ee

For the metric \eq{eq:genmetric} the directions of the coordinate line
$\theta,\phi$ and $\psi$ are eigen-vectors of $K_{ab}$ and the
corresponding eigen-values are $K_a$
\begin{equation}
K_{\psi}=\frac{R_{\x\phi\x\phi}}{fg}\hhh
K_{\phi}=\frac{R_{\x\psi\x\psi}}{fh}\hhh
K_{\x}=\frac{R_{\phi\psi\phi\psi}}{gh}\, .
\label{eq:K}
\end{equation}
These quantities are the curvatures of the 2D sections orthogonal to
$\psi$, $\phi$, and $\zeta$ lines, respectively. For brevity, we call
these 2D surfaces $\psi$-, $\phi$- and $\zeta$-sections.

For the unit sphere $S^3$, from \eq{eq:s33} one can easily see that
\be
K_{\psi}=K_{\phi}=K_{\theta}=1. 
\ee
However for the torus $S^2\times S^1$, from \eq{mt} we have 
\be
K_{\psi}=\frac{1}{b^2};\;\;\;  K_{\phi}=K_{\theta}=\frac{\cos\theta}{b(a+b\cos\theta)}.
\label{eq:ktori}
\ee
Thus we see that $K_{\psi}$ always remains positive, while $K_{\phi}$ and $K_{\hat\theta}$ 
is positive in the interval ($0\le\theta<\pi/2$). Thus the {\it equitorial plane} ($\theta=\pi/2$), 
divides the torus in two halves, one in which all the sectional curvatures are positive while 
the other has two of the sectional curvatures negative.  
In fact the surface $\theta=\pi/2$ is topologically $S^1\times S^1$ with the metric,
\be
ds^2=a^2d\psi^2+b^2d\phi^2
\ee   

The equations \eq{KK} imply
\begin{equation}
R^\x_\x=K_{\phi}+K_{\psi}\hhh
R^\phi_\phi=K_{\psi}+K_{\x}\hhh
R^\psi_\psi=K_{\x}+K_{\phi}\, .
\label{eq:Ricci}
\end{equation}
\begin{equation} 
R=2(K_{\x}+K_{\phi}+K_{\psi})\, .
\label{eq:Ricciscalar}
\end{equation}
From the above expression it is clear that $K_{\psi}<0$ if  $g'$ and
$\ln[fg/(g')^2]'$ have the opposite signs. Similarly $K_{\phi}<0$
imply $h'$ and $\ln[fh/(h')^2]'$ have opposite sign. For $K_{\x}<0$,
$g'$ and  $h'$ must have the same sign.

Let us consider now {\em Euler characteristics}  of the two dimensional sections of the horizon
surface. We denote by $\chi_a$ the Euler characteristic for the 2-surface $x^a=$const. 
By using the
{\em Gauss Bonnet} theorem we have,
\begin{equation}
2\pi\chi_a=\int\int_{{\cal M}}K_adA +\int_{\partial{\cal M}}k_gds\, .
\label{eq:Euler}
\end{equation}
Here $dA$ is the element of area on the surface and $k_g$ is the
geodesic curvature  on the boundary. If the surface has no boundary
or the boundary line is a geodesic,  then the last term vanishes. For
the metric (\ref{eq:genmetric}) simple calculations give
\ba
2\pi\chi_{\psi}&=&-\pi\left[\frac{g'}{\sqrt{fg}}\right]_{\x_{0}}^{\x_{1}}
+\int_{\partial{\cal M}}k_gds\, ,
\label{eq:Euler1}\\
2\pi\chi_{\phi}&=&-\pi\left[\frac{h'}{\sqrt{fh}}\right]_{\x_{0}}^{\x_{1}}
+\int_{\partial{\cal M}}k_gds\, ,
\label{eq:Euler2}\\
\chi_{\x}&=&0\, .
\label{eq:Euler3}
\ea
Thus we see that the Gaussian curvatures of sections, completely
describe the topology  and geometry of the 3-horizons.

\section{A 5D Rotating Black Hole with One Rotation Parameter}

\subsection{Volume and shape of the horizon surface}

For the 5 dimensional MP black hole with a single parameter of
rotation,  the induced metric on the horizon is  ~\cite{mp1},
\be
ds^2=r_0^2 ds_H^2\, ,
\ee
\begin{equation}
ds_H^2=f(\theta)d\theta^2+\frac{\sin^2\theta}{f(\theta)}d\phi^2
+(1-\al^2)\cos^2\theta d\psi^2\, .
\label{eq:mpmetric}
\end{equation}
Here $f(\theta)=(1-\al^2\sin^2\theta)$ and $r_0$ is length parameter
related with the mass $M$ of the black hole as 

\be
r_0^2=\frac{8\sqrt{\pi}GM}{3}
\ee 

The metric \eq{eq:mpmetric} is in the {\em Hopf coordinates} and
hence the co-ordinate $\theta$ varies from $0$ to $\pi/2$. The
rotation is along the  $\phi$ direction. The quantity $\al=|a|/r_0$
characterize the rapidity of  the rotation. It vanishes for a
non-rotating black hole and take the maximal value $\alpha=1$ for an
extremely rotating one. In what follows we put $r_0=1$, so that
$\alpha$ coincides with the rotation parameter. Different quantities
(such as lengths and curvature components) can be easily obtained
from the corresponding dimensionless expressions by using their
scaling properties.

\begin{figure}[htb!!]
\begin{center}
\includegraphics[height=6cm,width=5cm]{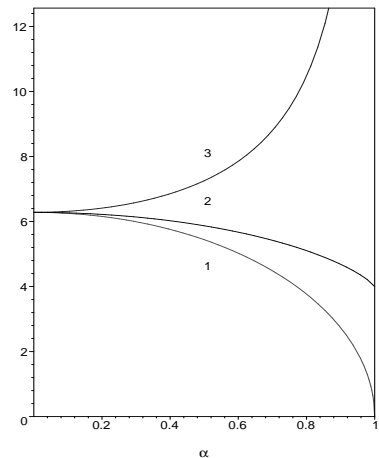}
\caption{Lengths $l_{\psi}$ (1), $l_{\theta}$ (2) and
$l_{\phi}$ (3) as the functions of the rotation parameter $\alpha$}
\label{f0}
\end{center}
\end{figure}

For $\alpha=0$ the horizon is a round sphere $S^3$ of the unit
radius. In the presence of rotation  this sphere is distorted.  Its
3-volume is $V_3=2\pi^2\sqrt{1-\alpha^2}$. In the limiting case of an
extremely rotating black hole, $\alpha=1$, $V_3$ vanishes.

The coordinate lines of $\phi$ and $\psi$ on this distorted sphere
are remain closed circles. The length of the circle corresponding to
the $\phi$-coordinate changes from 0 (at $\theta=0$) to its largest
value (at $\theta=\pi/2$)
\be
l_{\phi}={2\pi\over \sqrt{1-\alpha^2}}\, .
\ee
Similarly, the length of the circles connected with $\psi$-coordinate
changes from its maximal value (at $\theta=0$)
\be
l_{\psi}=2\pi \sqrt{1-\alpha^2}\, 
\ee 
to 0 at $\theta=\pi/2$. A line $\phi,\psi=$const on the disported
sphere is again a geodesic.
\be
l_{\theta}=4{\bf E}(\alpha)\, .
\ee

The lengths $l_{\psi}$, $l_{\theta}$  and $l_{\phi}$ as the functions
of the rotation parameter $\alpha$ are shown at figure~\ref{f0} by
the lines 1, 2, and 3, respectively. All these lines start at the
same point $(0,2\pi)$. In the limit of the extremely rotating black
hole ($\alpha=1$) the horizon volume vanishes,  $l_{\psi}=0$,
$l_{\theta}=4$, and $l_{\phi}$ infinitely grows.

\subsection{Gaussian curvature}

Calculations of the eigen-values $K_a$ of the Gaussian curvatures give
\begin{equation}
K_{\psi}=\frac{[1-\al^2(1+3\cos^2\theta)]}{f(\theta)^3}\hhh K_{\phi}=
K_{\theta}=\frac{1}{f(\theta)^2}\, .
\label{eq:Kmp}
\end{equation} 
From these relations it follows that the quantity $K_\psi$ is
negative in the vicinity of the `pole' $\theta=0$ for $1/2<\al<1$,
while the other two quantities, $K_\theta$ and $K_\phi$ are always
positive. This is similar to the 4D Kerr black hole  where the
Gaussian curvature of the two dimensional horizon becomes negative 
near the pole for $\al>1/2$. This is not surprising since the 2D
section $\psi=$const of the metric is isometric to the geometry of
the horizon surface of the Kerr black hole.

The  Ricci tensor and Ricciscalar for the metric on the surface of
the horizon of 5D black hole are 
\ba
R_\T^\T&=&R^\phi_\phi=\frac{2[1-\al^2(1+\cos^2\theta)]}{f(\theta)^3}\, , 
\label{eq:Rmp}\\
R_\psi^\psi&=&\frac{2}{f(\theta)^2}\hhh
R=\frac{2[3-\al^2(3+\cos^2\theta)]}{f(\theta)^3}\, .
\label{eq:Ricciscalarmp}
\ea
The components  of Ricci tensor $R_\T^\T$ and $R^\phi_\phi$ becomes
negative for certain values near the `pole' $\theta=0$, when
$\al>1/\sqrt{2}$, while the Ricci scaler is negative when 
$\al>\sqrt{3/4}$.

It is interesting to note that the surfaces of constant $\phi$ or
constant $\psi$ are  topologically disks with Euler Characteristics
equal to unity. The  boundary of these disks are on $\T=\pi/2$. It is
easy to check from equations  (\ref{eq:Euler1}) and (\ref{eq:Euler2})
that boundary terms of {\em Gauss Bonnet} equation  vanishes on this
boundary. This shows the boundary, which is the equatorial line on 
the deformed hemisphere, is a geodesic of the induced metric. Another
important point  is, while approaching the naked singularity limit
($\al=1$), the Gaussian curvatures of all  the three sections, as
well as the negative Ricciscalar blows up along the  `equator'
($\T=\pi/2$). This shows the extreme flattening of the horizon along 
equatorial plane, before the horizon shrinks to zero volume.

\section{Embedding}

\subsection{Embedding of the horizon in 5D pseudo-Euclidean space}

Let us discuss now the problem of the embedding of the horizon
surface of a rotating 5D black hole into a flat space. We start by
reminding that a similar problem for a 4D (Kerr) black hole was
considered long time ago by Smarr \cite{kerr1}. He showed that if the
rotation parameter of the Kerr metric $\alpha<1/2$, then 2D surface
of the horizon can be globally embedded in $\E^3$ as a rotation
surface. For $\alpha>1/2$ such an embedding is possible  if the
signature of the 3D flat space is $(-,+,+)$. In a recent paper
\cite{kerr2} there was constructed a global embedding of the horizon
of a rapidly rotating black hole into $\E^4$. 

Since the 3D surface of a rotation 5D black hole has 2
commuting orthogonal Killing vectors it is natural to consider
its embedding into the flat space which has at least two
independent orthogonal 2-planes of the rotation. In this case
the minimal number of the dimensions of the space of embedding
is 5. We write the metric in the form
\be
dS^2=\varepsilon dz^2+dx_1^2+dx_2^2+dy_1^2+dy_2^2\, ,
\ee
where $\varepsilon=\pm 1$. By introducing polar coordinates
$(\rho,\phi)$ and $(r,\psi)$ in the 2-planes $(x_1,x_2)$ and
$(y_1,y_2)$, respectively, we obtain
\be\n{dS}
dS^2=\varepsilon dz^2+d\rho^2+\rho^2 d\phi^2+dr^2+r^2 d\psi^2\, .
\ee
Using $\mu=\cos\theta$ as a new coordinate one can rewrite the metric on
the horizon \eq{eq:mpmetric} in the form
\ba
ds^2&=&f d\mu^2+\rho^2 d\phi^2+r^2 d\psi^2\, ,\n{ds}\\
f&=&{1-\alpha^2\mu^2\over 1-\mu^2}\hh 
\rho={\mu\over \sqrt{1-\alpha^2\mu^2}}\, ,\\
r&=&\sqrt{(1-\alpha^2)(1-\mu^2)}\, .\n{rr}
\ea
Assuming that $z$ is a function of $\mu$, and identifying $\rho$ and
$r$ in \eq{dS} with \eq{rr} one obtains the metric \eq{ds} provided the
function $z(\mu)$ obeys the equation
\be\n{dz}
\left({dz\over d\mu}\right)^2=\varepsilon \left[ f-\left({d\rho\over
d\mu}\right)^2-\left({dr\over d\mu}\right)^2\right]\, .
\ee
By substituting \eq{rr} into \eq{dz} one obtains
\be\n{dzz}
\left({dz\over d\mu}\right)^2=\varepsilon
{\alpha^2\mu^2(3\alpha^2\mu^2-\alpha^4\mu^4-3)\over (1-\alpha^2\mu^2)^3} \, .
\ee
It is easy to check that for $|\alpha|\le 1$ and $0\le\mu\le 1$ the
expression in the right hand side of \eq{dzz} always has the sign opposite
to the sign of $\varepsilon$.  Thus one must choose $\varepsilon=-1$ and one has
\be
z={1\over 2\alpha}\int_{\sqrt{1-\alpha^2\mu^2}}^1 {dy\over y^{3/2}}\sqrt{1+y+y^2}\, .
\ee
Let us emphasize that this result is valid both for the slowly and 
rapidly rotating black holes.

\subsection{Global embedding into $\E^6$}

\subsubsection{Construction of an embedding}

It is possible, however, to find a global isometric embedding of the 3-horizon of a rotating
black hole in a flat space with positive signature, if the number of dimensions is 6.  This
embedding is analogues to the one discussed  in \cite{kerr2} for the
rapidly rotating Kerr black hole.

Let us denote by  $X_i,  (i=1...6)$ the Cartesian co-ordinates in
$\E^6$. We write the embedding equations in the form
\begin{equation}
X_i=\frac{\eta(\theta)}{\rho_0}n^i(\pht)\, ,
\quad (i=1,2,3)\ ,
\label{eq:emeq1}
\end{equation}
\begin{equation}
X_4=\nu(\theta)\cos\psi\hhh 
X_5=\nu(\theta)\sin\psi \hhh
X_6=\chi(\theta)\, .
\label{eq:emeq2}
\end{equation}
Here the functions $n^i$ obey the condition
\begin{equation}
\sum_{i=1}^3 (n^i(\pht))^2=1\, .
\label{eq:cond}
\end{equation}
In other words, the 3D vector $n^i$ as a function of $\pht$ describes
a line on the unit round sphere  $S^2$. We require this line to be a
smooth closed loop (${\bf n}(0)={\bf n}(2\pi)$) without  self
interactions. We denote 
\begin{equation}
\rho(\pht)= \left[\sum_{i=1}^3 (n^i_{,\pht})^2\right]^{1/2}\, .
\label{eq:length1}
\end{equation}
Then $dl=\rho(\pht)d\pht$ is the line element along the loop.
The total length of the loop is 
\begin{equation}
l_0=2\pi\rho_0=\int_0^{2\pi}\rho(\pht)d\pht\, .
\label{eq:length2}
\end{equation}

We define a new co-ordinate $\ph$ as
\begin{equation}
\ph=\frac{1}{\rho_0}\int_0^{\pht}\rho(\pht)d\pht\, .
\end{equation}
It is a monotonic function of $\pht$ and  has the same period $2\pi$
as $\pht$. The induced metric for the embedded 3D surface defined by
(\ref{eq:emeq1}) and  (\ref{eq:emeq2}) becomes
\begin{equation}
ds^2=\left[\frac{{\eta_{,\theta}}^2}{\rho_0^2}+{\nu_{,\theta}}^2+{\chi_{,\theta}}^2\right]d\theta^2
+\eta^2d\phi^2+\nu^2d\psi^2\, .
\label{eq:embedmetric}
\end{equation}
Now comparing equations (\ref{eq:mpmetric}) and (\ref{eq:embedmetric}) 
we get
\ba
\eta(\theta)&=&\frac{\sin\theta}{\sqrt{1-\al^2\sin^2\theta}}\hhh
\nu(\theta)=\sqrt{1-\al^2}\cos\theta\, ,
\label{eq:embfunc1}\\
\chi(\theta)&=&\int_0^\theta\cos\theta\sqrt{\left[1-\frac{1}
{\rho_0^2(1-\al^2\sin^2\theta)^3}\right]}d\theta \, .
\label{eq:embfunc2}
\ea
We choose the functions $n^i$ in such a way that
\begin{equation}\n{cond}
\rho_0^2\ge\frac{1}{(1-\al^2)^3}\, ,
\end{equation}
so that the function $\chi(\theta)$ remains real valued for all
$\theta$ and hence  we can globally embed the horizon in $\E^6$.

\subsubsection{A special example}

\begin{figure}[tb!!]
\begin{center}
\includegraphics[height=4cm,width=4cm]{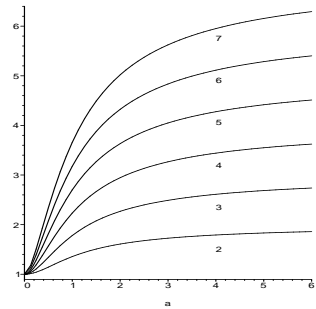}
\caption{$\rho_0$ as a function of $a$ for the values of $N$ from 2
(line 2) to 7 (line 7)}
\label{ff1}
\end{center}
\end{figure}

\begin{figure}[tb!!]
\begin{center}
\includegraphics[height=4cm,width=5cm]{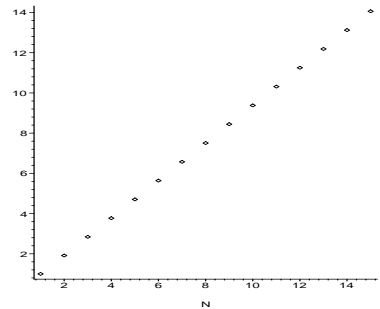}
\caption{$\rho_0$ as a function of $N$ for $a=10$.}
\label{f2}
\end{center}
\end{figure}

To give an explicit example of the above described embedding let us
put
\ba
n^1&=&\frac{\cos\pht}{F}\hhh
n^2=\frac{\sin\pht}{F}\hhh
n^2=\frac{a\sin(N\pht)}{F}\, ,
\label{eq:example}\\
F&=&\sqrt{1+a^2\sin^2(N\pht)}\, .
\ea
Here $N\ge 1$ is a positive integer. For this choice the value of
$\rho_0$ is
\begin{equation}
\rho_0=\frac{1}{2\pi}\int_0^{2\pi}\frac{\sqrt{a^2\cos^2(\pht)(N^2-1)+a^2+1}}
{1+a^2\sin^2(\pht)}d\pht\, .
\label{eq:rho0}
\end{equation}
For $N=1$ $\rho_0=1$.
For $N>1$ the above integral can be exactly evaluated to give 
\ba
\rho_0&=&\frac{2}{\pi}k_1\left[N^2\Pi(a^2k_1^2,ik_2)-(N^2-1)K(ik_2)\right]\,
,
\label{eq:rho1}\\
k_1&=&\frac{1}{\sqrt{1+a^2}}\hh
k_2=a\sqrt{\frac{N^2-1}{1+a^2}}\, .
\ea Here $K$ and $\Pi$ are elliptic integrals of first and third
kind, respectively. For a fixed value of $N$  $\rho_0$ is
monotonically growing function of $a$ (see Fig.~\ref{ff1}). For a
fixed value of $a$ the value of $\rho_0$ increases monotonically with
$N$ (see Fig.~\ref{f2}). The asymptotic form of $\rho_0$ for large
values of $a$ can be easily obtained as follows. Notice that for
large $a$ the denominator in the integral \eq{eq:rho0} is large
unless $\pht$ is close to $0$, $\pi$, or $2\pi$. Near these points
$\cos\pht$ can be approximated by 1, and the expression for $\rho_0$
takes the form
\be
\rho_0\approx {aN\over 2\pi} \int{d\pht \over 1+a^2\sin^2\pht}=
{aN\over \sqrt{1+a^2}}\, .
\ee
  
Using these properties of $\rho_0$, one can show that  for large
enough values of $N$ and $a$ the quantity $\rho_0$ can be made
arbitrary large, so that the condition \eq{cond} is satisfied and we have
the global embedding of the horizon surface for any $\al<1$.

\section{A 5D Rotating Black Ring}

\subsection{Horizon surface of a black ring}

Now we consider properties of horizon surfaces of  stationary black
strings in an asymptotically 5D flat spacetime ~\cite{br1}. In this paper 
we would only consider the {\em balanced} black ring in the sense that 
there are no angular deficit or
angle excess causing a conical singularity. The ring rotates along
the $S^1$ and this balances the gravitational self attraction.
The metric of the
rotating black ring is  ~\cite{br2,note} 
\begin{eqnarray}
ds^2 &=  -[F(x)/F(y)]\left[dt + r_0{
\displaystyle\frac{\sqrt{2}\nu}{\sqrt{1+\nu^2}}}(1+y)
d\tilde{\psi}\right]^2 \nonumber\\
 &+ {\displaystyle \frac{r_0^2}{(x-y)^2}}\left[-F(x)\left(G(y)d\tilde{\psi}^2
 +[F(y)/G(y)]dy^2\right)\right.  
\nonumber\\
&+F(y)^2 \left. \left([dx^2/G(x)]+[G(x)/F(x)]d\phi^2\right)\right]\, ,
\label{eq:blackring}
\end{eqnarray} 
where 
\be
F(\x)=1-\frac{2\nu}{1+\nu^2}\x\hhh
G(\x)=(1-\x^2)(1-\nu\x)\, .
\label{eq:fg}
\ee
The quantity $r_0$ is the radius scale of the ring.  The parameter
$\nu\in [0,1]$  determines the shape of the ring. The
coordinate $x$ changes in the the interval $-1\le x\le 1$ , while
$y^{-1}\in [-1,(2\nu)/(1+\nu^2)]$. The black ring is rotating in
$\tilde{\psi}$-direction. The positive `$y$' region is the ergosphere
of the rotating black ring while the negative `$y$' region is lies
outside the ergosphere with the spatial  infinity at $x=y=-1$.

The  metric \eq{eq:blackring}  has a co-ordinate singularity at 
$y=1/\nu$. However after the transformation 
\begin{eqnarray}
d\psi&=&d\tilde{\psi}+J(y)dy\, , \nonumber \\ 
dv&=&dt-r_0\frac{\sqrt(2)\nu}{\sqrt{1+\nu^2}}(1+y)J(y)dy\, ,
\label{eq:tran} 
\end{eqnarray} 
with $J(y)=\sqrt{-F(y)}/G(y)$,  the metric is regular at $y=1/\nu$.
In  these regular coordinates one can show that the surface $y=1/\nu$
is the  horizon.

The induced metric on
the horizon of a rotating black  ring is given by
\ba
ds^2&=&{r_0^2} ds_H^2\, ,\\
ds_H^2&=&\frac{p}{k(\T)}\left[\frac{d\T^2}{k(\T)^2}+\frac{\sin^2\T d\phi^2}
{l(\T)}\right]+ql(\T)d\psi^2\, ,
\label{eq:brmetric}\\
k(\T)&=&1+\nu\cos\T\hh l(\T)=1+\nu^2+2\nu\cos\T \, ,\\
p&=&{\nu^2(1-\nu^2)^2\over {1+\nu^2}}\hh q=2\frac{1+\nu}{(1-\nu){1+\nu^2}}\, .
\label{eq:AB}
\ea
In this metric the co-ordinates
$\phi$ and $\psi$ have a period of $2\pi$ and $\T\in[0,\pi]$. 
$\T=0$ is the axis pointing outwards ({\em i.e} increasing $S^1$  radius), 
while $\T=\pi$ points inwards.
The volume of the  horizon surface for the metric \eq{eq:brmetric} is
\be
V=8\sqrt{2}\pi^2 \nu^2\sqrt{1-\nu}\left[\frac{\sqrt{1+\nu}}{\sqrt{1+\nu^2}}\right]^3\, .
\ee

\subsection{Gaussian curvature}

The metric \eq{eq:brmetric} is of the form \eq{eq:genmetric}, so that one can
apply to it the results of Section~III. For example,
its Gaussian curvatures has the following eigen-values 
($i=\psi,\phi,\theta$)
\be
K_i=\frac{k(\T)^2 F_i(\cos\T)}{2\nu(1-\nu^2)^2(1+\nu^2)l(\T)^2}\,,
\ee
where the functions $F_i=F_i(\x)$ are defined as follows
\ba
F_{\psi}&=&\nu(3+\nu^2)\x^2+2(\nu^4+\nu^2+2)\x+2\nu^{-1}-\nu+3\nu^3\, ,\non
\label{eq:F1}\\
F_{\phi}&=&8\nu^2\x^3+\nu(5\nu^2+7)\x^2+2(1-\nu^2)\x-\nu(3\nu^2+1)\, ,\non
\label{eq:F2}\\
F_{\theta}&=&\nu^3\x^2+(6\nu^2+3\nu+2)\x+\nu(3+\nu^2)\, .
\label{eq:F3}
\ea

From the above equations it is clear that for any value of $\nu$ and
$\theta$  the Gaussian curvature of the $\psi$-sections, ({\em i.e}
$K_\psi$),  always remains positive. This is absolutely similar to the flat torus case
as described by \eq{eq:ktori}. The sign of Gaussian curvatures
for other two sections,  $K_\phi$ and $K_\theta$, depend on
the values of $\nu$ and $\theta$. For example, for $\T=0$, both
these curvatures are positive for all values of $\nu$.  But as we
increase $\theta$, $K_\phi$ becomes negative for higher values of 
$\nu$ and ultimately when $\T=\pi/2$ it becomes negative for all
$\nu$ and continue  to be negative till $\T=\pi$. On the other hand $K_\theta$ 
remains positive and grows with $\nu$ 
for all $\nu$ till $\T=\pi/2$, then starts becoming negative for higher values 
of $\nu$ and ultimately becomes negative for all $\nu$ at $\T=\pi$.

\begin{figure}[tb!!]
\begin{center}
\includegraphics[width=5cm]{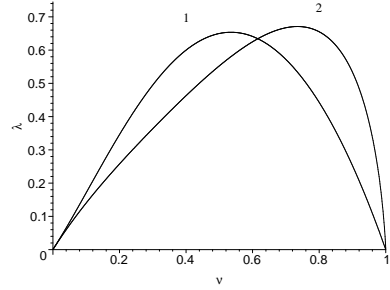}
\caption{$\lambda_{\theta 1}$ and $\lambda_{\theta 2}$ as a function of $\nu$.}
\label{f8}
\end{center}
\end{figure}

\begin{figure}[tb!!]
\begin{center}
\includegraphics[width=5cm,height=4cm]{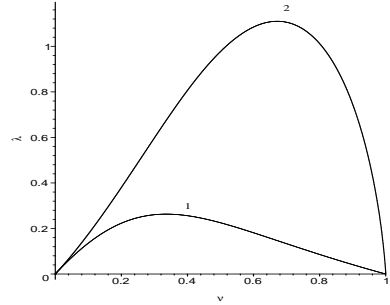}
\caption{$\lambda_{\phi 1}$ and $\lambda_{\phi 2}$ as a function of $\nu$.}
\label{f9}
\end{center}
\end{figure}

Let us emphasize that, because of the distortions due to the rotation, 
both $K_\phi$ and $K_\theta$ do not become negative at the same value of $\theta$ 
as it was for the flat torus case. To get an invariant measure of distortion produced 
due to rotation, let us define two invariant lengths in the following way.
Let $\theta=\theta_i$ ($i=\theta,\phi$), be the point where $F_i(\cos\theta)$ 
vanishes. Then the two invariant lenghts are,
\be
\lambda_{i1}=2\sqrt{p}\int_0^{\theta_i}\frac{d\theta}{\sqrt{k(\theta)}^3},\;\;
\lambda_{i2}=2\sqrt{p}\int_{\theta_i}^\pi\frac{d\theta}{\sqrt{k(\theta)}^3}.
\label{eq:invlength}
\ee
It is easy to check from \eq{mt} that in the case of flat tori we have 
$\lambda_{i1}=\lambda_{i2}$. However for the rotating black rings they are 
different functions of the parameter $\nu$. Figures ~\ref{f8} and ~\ref{f9} 
shows the invariant lenghts 
for ($i=\theta,\phi$) as function of $\nu$ respectively. We see that for $i=\theta$, 
$\lambda_{i2}<\lambda_{i1}$ for small 
$\nu$. However as we increase $\nu$ the difference between them 
reduces and ultimately at 
$\nu\approx0.615$, $\lambda_{i2}$ overtakes $\lambda_{i1}$. 
Whereas for $i=\phi$, $\lambda_{i2}$ is
always greater than $\lambda_{i1}$.

It is evident that both $\psi$- and $\phi$-sections are closed and do
not have a  boundary. Calculating the {\em Euler numbers} for these
surface we get
\begin{equation}
\chi_{\psi}=2\hh \chi_\phi=0\, .
\end{equation}
This shows that the $\psi$-section is a deformed 2-sphere with
positive  Gaussian curvature. Its rotation in the $\psi$ direction 
generates the horizon surface of the rotating black ring.

\subsection{Kaluza-Klein Reduction of Rotating Black Ring}

The absence of the cone-like singularities in the black ring solution
\eq{eq:blackring} is a consequence of the exact balance between the
gravitational attraction and the centrifugal forces generated by the
ring's rotation. We discuss the effects connected with the ring
rotation from a slightly different point of view. Let us write the
metric \eq{eq:blackring} in the following Kaluza-Klein form (see e.g.
\cite{kk3})
\begin{equation}
ds_5^2=\Phi^{-\frac{1}{3}}\left[h_{\alpha\beta}dx^\alpha dx^\beta
+\Phi (A_t dt +d\phi)^2\right]\, .
\label{eq:kk5}
\end{equation}

The 4D reduced {\it Pauli} metric in this space is ($a,b=0,\ldots,3$)
\begin{equation}
ds_4^2= h_{ab}dx^a dx^a
=\Phi^{\frac{1}{3}}\left[g_{ab}dx^a dx^b
-\Phi A_t^2dt^2\right]\, .
\label{eq:kk4}
\end{equation}
Here $g_{ab}$ is the four dimensional metric  on the $\tpsi$-section
of the 5D black ring. By the comparison of \eq{eq:kk5} and
\eq{eq:kk4} one has
\ba
\Phi^{\frac{2}{3}}&=&\xi_{\tpsi}^2=-\frac{F(x)}{F(y)}L(x,y)\, ,\\
A_t&=&\frac{(\xi_t.\xi_{\tpsi})}{\xi_{\tpsi}^2}
=\frac{\sqrt{2}\nu}{\sqrt{1+\nu^2}}\frac{(1+y)}{L(x,y)}\, ,\\
L(x,y)&=&\left[\frac{2\nu^2}{1+\nu^2}(1+y)^2+\frac{F(y)G(y)}{(x-y)^2}\right]\,,
\label{eq:phi}
\ea
where  $\xi_t=\partial_t$, $\xi_{{\tpsi}}=\partial_{\tpsi}$ and
$\xi_{\phi}=\partial_{\phi}$ are the Killing vectors of \eq{eq:kk5}.
The quantities $\ln(\Phi)$ and $A_t$ can be interpreted as a dilaton
field and a `electromagnetic potential' in the 4D spacetime. 

The horizon for the 4D metric \eq{eq:kk4} is defined by the condition
\begin{equation}
h_{tt}=\xi_t^2-({\xi_{\tpsi}^2})^{-1}(\xi_t.\xi_{\tpsi})^2=0\, .
\end{equation}
It is easy to show that this condition is equivalent to the condition
defining the horizon of the 5D metric. Thus both horizons are
located at $y=1/\nu$. 

\begin{figure}[tb!!]
\begin{center}
\includegraphics[width=5cm, height=3cm]{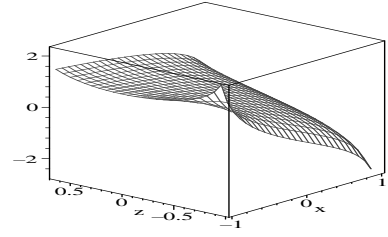}
\caption{$\ln(\Phi)$ as a function of $x$ and $z=1/y$ for  
$\nu=2/3$.}
\label{f5}
\end{center}
\end{figure}

To summarize, the 4D metric \eq{eq:kk4} obtained after the reduction
describes a static 4D black hole in the presence of an external
dilaton and `electromagnetic' field. The dilaton field $\ln \Phi$ (as
well as the metric \eq{eq:kk4}) has a singularity at the points where
$\xi_{{\tpsi}}^2$ either vanish (at the axis of symmetry,
$x=1, y=-1$) or infinitely grows (at the spatial
infinity,  $x=y=-1$) (see Fig.~\ref{f5}). Outside these regions the dilaton field is
regular everywhere including the horizon where it takes the value
\begin{equation}
\Phi_{H}=\left[\frac{2}{1+\nu^2}
\left(\frac{1+\nu}{1-\nu}\right)l(\theta)\right]^{\frac{3}{2}}
\end{equation}

The `electromagnetic field strength', which has  non zero components 
\begin{equation}
F_{tx}=-A_{t,x}\hh F_{ty}=-A_{t,y}\, ,
\end{equation}
is regular everywhere and vanishes at the spatial infinity. However
the $F^2=F_{ab}F^{ab}$ invariant is well defined throughout the space
time but drops  (towards negative infinity) at the axis of symmetry.
Figure 8 illustrates the behavior of the invariant $F^2$  for
$\nu=2/3$.

\begin{figure}[tb!!]
\begin{center}
\includegraphics[width=5cm, height=3cm]{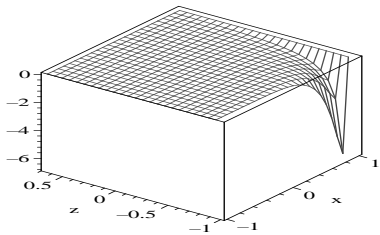}
\caption{The invariant  $F^2$ as a function of $x$ and $z=1/y$ for  $\nu=2/3$. It can be seen 
that it is well defined everywhere but  drops 
(towards negative infinity) at the axis of symmetry, $x=1, y=-1$.}
\label{f6}
\end{center}
\end{figure}

The metric $ds_1^2$ on 2D horizon surface for the reduced metric \eq{eq:kk4} is
conformal to the metric $ds_0^2$ of 2D section $\tpsi=$const of the black
ring horizon, \eq{eq:brmetric}. These metrics are of the form
($k=k(\theta)$, $l=l(\theta)$)
\be
ds_{\epsilon}^2=\Phi_H^{\epsilon/3}\frac{p}{k}\left[\frac{d\T^2}{k^2}
+\frac{\sin^2\T d\phi^2}{l}\right]\hh \epsilon=0,1\, .
\ee
Both metrics, $ds_{\epsilon}^2$
can  be embedded in  $\E^3$ as rotation surfaces.  The embedding
equations are
\begin{equation}
X_1=m\beta\cos\phi;\;
X_2=m\beta\sin\phi;\;
X_3=m\gamma\, ,
\label{eq:brembed1}
\end{equation}
where, 
\ba
\beta&=&k^{-1/2}l^{-1/2+\epsilon/4}\sin\theta \, ,\nonumber\\
\gamma&=&\int_0^\theta (k^{-3}l^{\epsilon/2}-\beta_{,\theta}^2)^{1/2}\,d\theta,\nonumber\\
m&=&\sqrt{p}\left[\frac{2}{1+\nu^2}({1+\nu\over 1-\nu})\right]^{\epsilon}\, .
\label{eq:brembed4}
\ea

\begin{figure}[tb!!]
\begin{center}
\includegraphics[width=3cm]{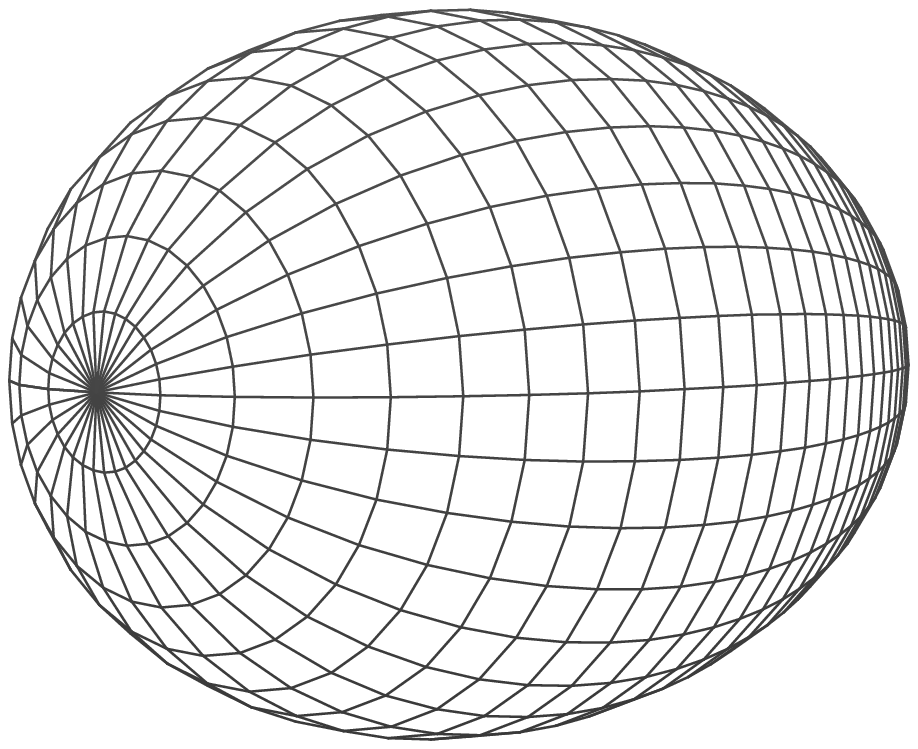}\hspace{0.5cm}
\includegraphics[width=2.5cm]{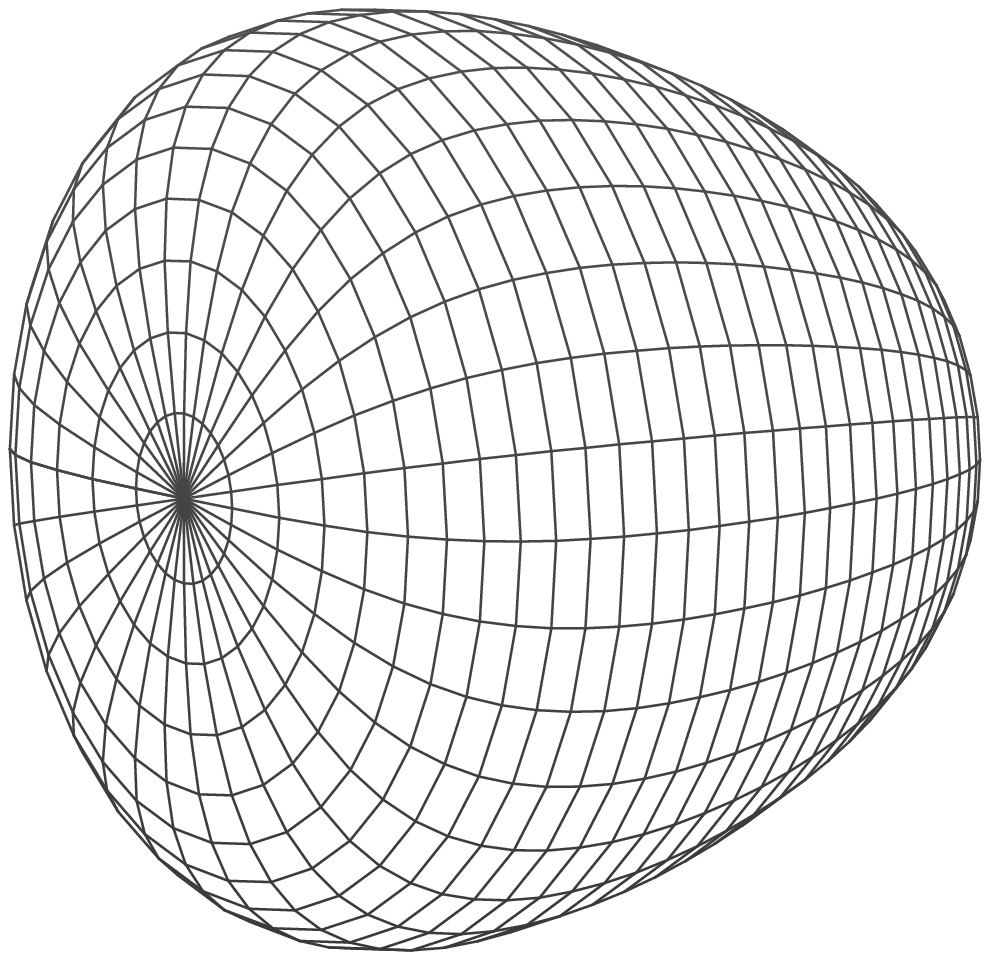}
\caption{The embedding diagrams for the metrics $ds_0^2$ (to the left)
and $ds_1^2$ (to the right) for $\nu=2/3$.}
\label{f7}
\end{center}
\end{figure}

The embedding diagrams for the metrics $ds_0^2$ and $ds_1^2$ are
shown in figure 9 by the left and right plots, respectively. Both
rotation surfaces are  deformed spheres. The surface with the
geometry $ds_1^2$ is more  flattened at poles.


\section{Discussion}

In this paper we discussed and analyzed  the surface geometry  of
five dimensional black hole and black rings with one parameter of
rotation.  We found that the sectional Gaussian curvature and  the
Ricci scalar of the horizon surface of the 5D rotating black hole are
negative if the rotation parameter is greater than some critical
value, similarly to the case of 4D the Kerr black hole. However there
is an important  difference between the embeddings  of the   horizon
surfaces of 5D and 4D black holes in the flat space.  As was shown in
~\cite{kerr1}, a rotating 2-horizon can  be embedded as a surface of
rotation in a 3-dimensional  Euclidean space only when the rotation
parameter is less than the critical value.  For the `super-critical'
rotation the global embedding is possible either in 3D flat space
with the signature if the metric $(-,+,+)$ \cite{kerr1} or in $\E^4$
with the positive signature \cite{kerr2}. For the 5D black hole for
any value of its rotation parameter the the horizon surface cannot be
embedded in 5D Euclidean space as a surface of rotation. Such an
embedding requires that the signature of the flat 5D space is
$(-,+,+,+,+)$. However we found a global embedding of this surface in
6D Euclidean space. 

We  calculated the surface invariants for the rotating black ring and
analyzed the effect of rotation on these  invariants. Finally we
considered the Kaluza-Klein reduction of rotating black ring which
maps its metric onto the metric of 4D black hole in the presence of
external dilaton and `electromagnetic' fields. Under this map, the
horizon of the 5D black ring transforms into the horizon of 4D black
hole. The `reduced' black hole is static and axisymmetric.  Distorted
black holes in the Einstein-Maxwell-dilaton gravity were discussed in
\cite{Ya}. This paper generalizes the well known results of
\cite{GH,Ch} for vacuum distorted black holes. It would be
interesting to compare the `reduced' distorted black hole discussed
in this paper with solutions presented in \cite{Ya}.

\noindent  

\section*{Acknowledgments}  
\noindent  
The research was supported in part by the Natural
Sciences and Engineering Research Council of Canada and by the Killam
Trust.



\begin{thebibliography}{99}
\bibitem{note} It should be emphasized that different publications used different
form the the black ring metric (see e.g. ~\cite{br2}-\cite{br5}).
\end{thebibliography}
\end{document}